\documentclass[prl,reprint,aps,showpacs,amsmath,amssymb,floatfix]{revtex4-2}
\usepackage{tikz}
\usetikzlibrary{arrows.meta,calc,shadings}
\usepackage{physics,mathtools}
\usepackage{pgfplots}
\usepgfplotslibrary{groupplots}
\pgfplotsset{compat=newest}
\usepackage{url}
\usepackage{amsmath}
\usepackage{amssymb}
\usepackage{bm}
\usepackage{graphicx}
\usepackage{hyperref}
\usepackage{dcolumn}
\usepackage{bm}
\usepackage{xcolor}
\newif\ifshowchanges
\showchangestrue

\usepackage{siunitx}
\usepackage[caption=false]{subfig}
\usepackage{multirow}
\usepackage{tabularx}

\hypersetup{
	colorlinks=true,       
	linkcolor=blue,        
	citecolor=blue,        
	filecolor=magenta,     
	urlcolor=blue         
}

\graphicspath{ {images/} }
\bibpunct{[}{]}{,}{n}{}{}

\def\tanh{\text{tanh}}

\begin{document}
\title{Thermomagnonic Torques in Insulating Altermagnets}
\author{Edward Schwartz}
\affiliation{Department of Physics and Astronomy and Nebraska Center for Materials and Nanoscience, University of Nebraska, Lincoln, Nebraska 68588, 
USA}
\author{Hamed Vakili}
\affiliation{Department of Physics and Astronomy and Nebraska Center for Materials and Nanoscience, University of Nebraska, Lincoln, Nebraska 68588, USA}
\author{Alexey A. Kovalev}
\affiliation{Department of Physics and Astronomy and Nebraska Center for Materials and Nanoscience, University of Nebraska, Lincoln, Nebraska 68588, USA}

\begin{abstract}We develop a symmetry-controlled theory of anisotropic thermomagnonic torques in insulating altermagnets. Starting from the stochastic LLG equations, we identify a spin-splitter magnonic torque associated with thermally generated, sublattice-odd spin currents and an anisotropic entropic torque dictated by crystal symmetry. These torques produce anisotropic magnetic-texture responses to temperature gradients. In particular, thermally generated spin currents induce domain-wall precession, which reduces domain-wall velocities for selected gradient directions. We also predict an anisotropic skyrmion Hall response, with symmetry-selected directions enabling fast skyrmion motion with strongly suppressed transverse deflection. Our results reveal experimentally testable symmetry fingerprints of insulating altermagnets and extend more broadly to anisotropic magnets with exchange-driven magnon spin splitting.
\end{abstract}
\maketitle

Materials with nonrelativistic spin splitting, such as altermagnets, promise exceptional spintronic properties arising from their spin-dependent electronic structure~\cite{Krempasky2024Altermagnetic,
Zhu2024Plaid,
Yamada2025PWaveMagnet,
Song2025ElectricalSwitching,
Hayami2019MomentumDependent,
Smejkal2020CrystalTimeReversal,
Naka2019SpinCurrentOrganic,
Sandratskii1981MnTe,
Mazin2021FeSb2,
Yuan2021LowZ,
Yuan2020GiantMomentumDependent,
Ma2021Multifunctional,
Smejkal2022Emerging,
Smejkal2022BeyondConventional,
Mazin2022Editorial,
Liu2025DifferentFacets,
Guo2023SpinSplit}. They can be characterized by spin symmetries that allow spin splitting of electronic bands, with $d$-, $g$-, or $i$-wave character depending on the material. Altermagnets combine useful properties of antiferromagnets (e.g., fast dynamics) with ferromagnet-like spin splitting and time-reversal-odd responses, enabling the crystal anomalous Hall effect~\cite{Smejkal2020CrystalTimeReversal,Smejkal2022NatRevMater,PhysRevLett.133.086503}, the spin-splitter effect~\cite{Naka2019SpinCurrentOrganic,PhysRevLett.126.127701,PhysRevLett.128.197202,Fu2025new}, anisotropic magnon spectra with lifted degeneracy~\cite{PhysRevLett.131.256703,PhysRevLett.131.186702}, and spin currents induced by temperature gradients~\cite{PhysRevB.108.L180401,PhysRevB.110.094427,PhysRevB.110.144421,Wu2025ChinPhysLett,GalindezRuales2025arXiv}. These and other useful features of altermagnets have driven rapid progress in recent years~\cite{Jungwirth2025AltermagneticSpintronics,Fukaya2025}.

The dynamics of magnetic textures such as domain walls~\cite{Gomonay2024,zn8d-ft9b} and skyrmions can be controlled by charge currents in altermagnets~\cite{PhysRevLett.134.176401}. A potentially more energy-efficient approach is to use temperature gradients, as explored in spin caloritronics~\cite{Kikkawa2023AnnuRevConMatPhys}. Altermagnets offer advantages for spin caloritronics because spin currents can be generated via the spin Seebeck effect without involving charge currents~\cite{PhysRevB.108.L180401,PhysRevB.110.094427,PhysRevB.110.144421,Wu2025ChinPhysLett}. The angular momentum carried by magnons can be exploited to control magnetic textures, as it can exert spin-transfer torques on them~\cite{PhysRevLett.107.177207,Kovalev2012,PhysRevB.86.054445,PhysRevLett.107.027205}. A second mechanism for controlling magnetic textures is related to entropic forces that arise from gradients in temperature-dependent magnetic parameters, such as the exchange stiffness~\cite{PhysRevLett.113.097201,PhysRevLett.117.107201,Kovalev2012,PhysRevB.92.020410}.

\begin{figure}
\centering
\includegraphics{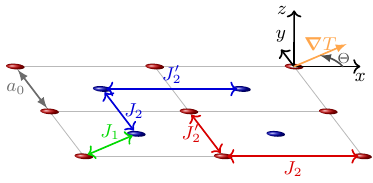}
\caption{A minimal model of a two-sublattice altermagnet. The two magnetic sublattices are shown as red and blue dots. The exchange interactions are indicated by double arrows. The unit cell, of area \(v=a_0^2\), contains two magnetic sites. The temperature gradient $\boldsymbol\nabla T$ is applied at an angle $\Theta$ with respect to the $x$ axis.}
\label{fig:altermagnet}
\end{figure}

In this paper, we develop a theory of thermomagnonic torques in insulating altermagnets. We apply our approach to a minimal model of a $d$-wave altermagnet~\cite{PhysRevB.108.224421} shown in Fig.~\ref{fig:altermagnet}; however, our theory can be readily generalized to other lattice models. We predict the existence of a spin-splitter thermomagnonic torque and show how it affects temperature-gradient-induced dynamics of magnetic textures. We also predict an anisotropic entropic torque that arises from the reduced symmetry of an altermagnet. We apply our theory to the dynamics of magnetic textures such as skyrmions and domain walls. The entropic torque drives domain walls and skyrmions toward the hot region. The magnonic spin-splitter torque induces domain wall precession, thereby slowing motion for certain crystal alignments. We also predict a temperature-gradient-driven anisotropic skyrmion Hall effect arising from the magnonic spin-splitter torque, which enables fast skyrmion motion along a nanotrack in response to thermal gradients for certain crystal alignments.

\textit{Thermomagnonic torques.} -- 
We consider the stochastic Landau-Lifshitz-Gilbert (LLG) equation written on a lattice,
\begin{equation}
s \bigl(1 + \alpha \mathbf{S}_\mathbf{r} \times \bigr) \, \partial_t \mathbf{S}_\mathbf{r}
   =\frac{2}{v} \mathbf{S}_\mathbf{r} \times \mathbf{h}_\mathbf{r} \, ,\label{eq:LLG}
\end{equation}
where $s=2s_\text{sub}$ with $s_\text{sub}$ being the spin density for one sublattice, \(v\) is the area/volume associated with a unit cell, $\mathbf{r}$ denotes a site belonging to one of the two sublattices, $\alpha$ stands for the Gilbert damping. Each unit vector represents angular momentum $\hbar S=s_\text{sub}v=sv/2$, so $S=sv/(2\hbar)$. The effective field $\mathbf{h}_\mathbf{r}$ is defined as  
$\mathbf{h}_\mathbf{r} = -\frac{\partial \mathcal{H}}{\partial \mathbf{S}_\mathbf{r}} + \boldsymbol{\zeta}_\mathbf{r}$. 
The stochastic field $\boldsymbol{\zeta}_\mathbf{r}$ is defined using the relations
\begin{equation}\label{eq:corr}
\big\langle \zeta^\alpha_{\mathbf{r}_1}(\omega) \,
\zeta_{\mathbf{r}_2}^\beta(\omega')^* \big\rangle
= \frac{2\pi \, \delta_{\alpha\beta} \, \alpha s_\text{sub} v \, \hbar \omega}
       {\tanh[\hbar \omega / (2k_B T)]} \,
  \delta_{\mathbf{r}_1\mathbf{r}_2} \,
  \delta(\omega - \omega') \, .
\end{equation} 
We represent the unit vector $\mathbf{S}_\mathbf{r}$ entering Eq.~(\ref{eq:LLG}) via two orthogonal components:
\(
\mathbf{S}_\mathbf{r} = \sqrt{1 - (S^{^\perp}_\mathbf{r})^2}\,\mathbf{S}_\mathbf{r}^{(0)} + \mathbf{S}_\mathbf{r}^{^\perp}.
\)
The slowly varying component is represented by the unit vector $\mathbf{S}_\mathbf{r}^{(0)}$, which corresponds to a smooth magnetic texture, while 
the fast component $\mathbf{S}_\mathbf{r}^{^\perp}$ stands for a small deviation  
from $\mathbf{S}_\mathbf{r}^{(0)}$ caused by thermal field. 

To identify thermomagnonic torques in Eq.~\eqref{eq:LLG}, we concentrate on the transverse dynamics and split the fast and slow components in the LLG equation
\begin{align}
    s \bigl( 1 + \alpha \chi_\mathbf{r} \mathbf{S}_\mathbf{r}^{(0)} \times \bigr) \partial_t \mathbf{S}^{(0)}_\mathbf{r} 
=&\frac{2}{v} \bigg[ \mathbf{S}_\mathbf{r}^{(0)} \times \mathbf{h}_\mathbf{r}^{(0)} \chi_\mathbf{r}\bigg. \\ \label{eq:torque1}
-& \sum_{\boldsymbol\delta}  J_{\boldsymbol\delta} \mathbf{S}_\mathbf{r}^{(0)} \times \mathbf{S}_{\mathbf{r}+\boldsymbol\delta}^{(0)}  (\chi_{\mathbf{r}+\boldsymbol\delta}-\chi_\mathbf{r}) \\ \label{eq:torque2}
-&\bigg. \sum_{\boldsymbol\delta}  J_{\boldsymbol\delta} \langle \mathbf{S}^{^\perp}_\mathbf{r} \times \mathbf{S}_{\mathbf{r}+\boldsymbol\delta}^{^\perp} \rangle \bigg] \, ,
\end{align}
where we use the exchange magnon approximation. Above, \(
\mathbf{h}_\mathbf{r}^{(0)}=-\frac{\partial \mathcal{H}}{\partial \mathbf{S}^{(0)}_\mathbf{r}} \)
is the effective field for \(\mathbf{S}_\mathbf{r}^{(0)}\) calculated at zero temperature, and $\chi_\mathbf{r}=\langle\sqrt{1 - (S^{^\perp}_\mathbf{r})^2}\rangle$ 
describes the shortening of the spin field. 

We assume that the field \(\mathbf{S}_\mathbf{r}^{(0)}\) varies slowly and can be represented by the N\'eel field 
\(\mathbf{n}=(\mathbf{S}_{A}^{(0)}-\mathbf{S}_{B}^{(0)})/2\) where the index \(A\) (\(B\)) stands for one of two sublattices. We also define the small canting field $\mathbf m=(\mathbf S_A^{(0)}+\mathbf S_B^{(0)})/2$. The torques below are contributions to $\partial_t\mathbf m$ and $\partial_t\mathbf n$.
After rewriting the magnetization-shortening term
\(
-\sum_{\boldsymbol\delta} J_{\boldsymbol\delta}\,
\mathbf{S}_{\mathbf r}^{(0)} \times \mathbf{S}_{\mathbf r+\boldsymbol\delta}^{(0)}
(\chi_{\mathbf r+\boldsymbol\delta}-\chi_{\mathbf r})
\)
in terms of the N\'eel field \(\mathbf n\), we obtain the torque densities
\begin{align}
    \boldsymbol \tau_m^0
    &=-2\,\mathbf n\times (\beta\mathbf u_0\cdot\boldsymbol\partial)\mathbf n,
    \label{eq:entropic} \\
    \boldsymbol \tau_n^0
    &=-2\,\mathbf n\times (\beta^\prime\mathbf u_0^\prime\cdot\boldsymbol\partial)\mathbf n,
    \label{eq:entropic3}
\end{align}
with
\(
(\beta u_0)_i
=-(A^A_{0,ij}+A^B_{0,ij})\,\partial_j\chi/(2s)\),
\((\beta^\prime u_0^\prime)_i
=-(A^A_{0,ij}-A^B_{0,ij})\,\partial_j\chi/(2s)
\).
Here, the zero-temperature exchange stiffness for sublattice \(A\) or \(B\) is
\begin{equation}
    A_{0,ij}^{A/B}
    =-\frac{1}{v}\sum_{\boldsymbol \delta}\delta_i\delta_j\,J_{\boldsymbol\delta}
    \left(\mathbf{S}_\mathbf r^{(0)}\cdot \mathbf{S}_{\mathbf{r}+\boldsymbol\delta}^{(0)}\right)_{\rm col},
    \label{eq:exch}
\end{equation}
where the subscript ``col'' indicates that the scalar product is evaluated in the collinear reference state.

\begin{figure}
\centering
\includegraphics{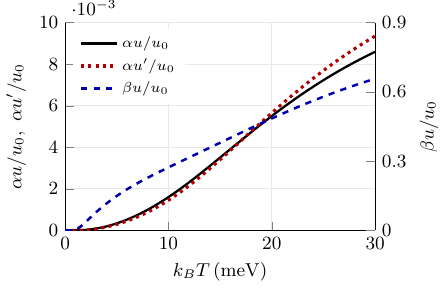}
\caption{Entropic, conventional magnonic, and spin-splitter torque coefficients for $\boldsymbol\nabla T\parallel+\hat{\mathbf x}$. The left axis shows $\alpha u/u_0$ (solid black) and $\alpha u^\prime/u_0$ (dotted red); the right axis shows $\beta u/u_0$ (dashed blue). The sign convention is specified in the text. Here \(s=\hbar/v\), \(u_0=(k_B/s)|\boldsymbol\nabla T|\), \(J_1=11.1~\mathrm{meV}\), \(J_2=-0.28~\mathrm{meV}\), \(J_2^\prime=-3.48~\mathrm{meV}\), and \(K=0.047~\mathrm{meV}\)~\cite{Gomonay2024}.}
\label{fig:torques}
\end{figure}

Next, we rewrite the fluctuation term
\(
-\sum_{\boldsymbol\delta} J_{\boldsymbol\delta}
\langle \mathbf{S}^{\perp}_\mathbf r \times \mathbf{S}_{\mathbf r+\boldsymbol\delta}^{\perp} \rangle
\)
by separating it into components transverse and longitudinal with respect to
\(\mathbf{S}_\mathbf r^{(0)} \times \mathbf{S}_{\mathbf r+\boldsymbol\delta}^{(0)}\).
The transverse contribution can be rewritten as the magnonic spin-transfer torque~\cite{Kovalev2012}
\begin{align}
\boldsymbol\tau_m^\text{st}&=\left(\mathbf{u}^\prime\cdot\boldsymbol\partial\right)\mathbf{n},
\label{eq:st1}\\
\boldsymbol\tau_n^\text{st}&=-\left(\mathbf{u}\cdot\boldsymbol\partial\right)\mathbf{n},
\label{eq:st2}
\end{align}
where
\(
\mathbf{u}=-(\mathbf{j}_{(1)}+\mathbf{j}_{(2)})/(2s)\),
\(
\mathbf{u}^\prime=(\mathbf{j}_{(1)}-\mathbf{j}_{(2)})/(2s)\),
and the spin current density is calculated for the first, \(\mathbf{j}_{(1)}\), or the second, \(\mathbf{j}_{(2)}\), sublattice site~\cite{PhysRevB.89.241101,PhysRevB.92.020410}. It is given by
\begin{equation}
\mathbf{j}_\mathbf{r}
=-\frac{1}{v}\sum_{\boldsymbol\delta}\boldsymbol\delta\,J_{\boldsymbol\delta}
\left(
\mathbf{S}_\mathbf r^{(0)}\cdot
\langle \mathbf{S}^{\perp}_\mathbf r \times \mathbf{S}^{\perp}_{\mathbf r+\boldsymbol\delta} \rangle
\right),
\label{eq:current}
\end{equation}
and is evaluated in the absence of a magnetic texture. The minus sign in the definition of $\mathbf u$ corresponds to the adiabatic-torque convention used below; $\mathbf u$ points along $\boldsymbol\nabla T$ for the parameters in Fig.~\ref{fig:torques}.

Assuming circular magnons, the remaining longitudinal contribution of the fluctuation term can be combined with
Eqs.~\eqref{eq:entropic} and \eqref{eq:entropic3}, but it enters with one-half of the weight of the bond-correlator contribution to the full finite-temperature exchange stiffness. It is therefore useful to distinguish between the full finite-temperature stiffness 
\begin{equation}
    A_{ij}^{A/B}
    =-\frac{1}{v}\sum_{\boldsymbol \delta}\delta_i\delta_j\,J_{\boldsymbol\delta}
    \left\langle\mathbf{S}_\mathbf r\cdot \mathbf{S}_{\mathbf{r}+\boldsymbol\delta}\right\rangle,
    \label{eq:exch-1}
\end{equation}
and the effective stiffness that governs the thermomagnonic torque.
\begin{equation}
    \widetilde A_{ij}^{A/B}
    \equiv
    A_{ij}^{A/B}+\frac{1}{2v}\sum_{\boldsymbol \delta}\delta_i\delta_j\,J_{\boldsymbol\delta}\,
    \left\langle \mathbf{S}_\mathbf r^{\perp}\cdot \mathbf{S}_{\mathbf{r}+\boldsymbol\delta}^{\perp}\right\rangle.
    \label{eq:exch-torque}
\end{equation}
Using \(\widetilde A_{ij}^{A/B}\), the total entropic torques read
\begin{align}
    \boldsymbol \tau_m^\text{tot}
    &=-\mathbf{n}\times (\beta\mathbf{u}\cdot\boldsymbol\partial)\mathbf{n},
    \label{eq:entropic1} \\
    \boldsymbol \tau_n^\text{tot}
    &=-\mathbf{n}\times (\beta^\prime\mathbf{u}^\prime\cdot\boldsymbol\partial)\mathbf{n},
    \label{eq:entropic2}
\end{align}
with
\(
(\beta u)_i
=-(\partial_j \widetilde A^A_{ij}+\partial_j \widetilde A^B_{ij})/(2s)\),
\((\beta^\prime u^\prime)_i
=-(\partial_j \widetilde A^A_{ij}-\partial_j \widetilde A^B_{ij})/(2s)\).
Thus, Eq.~\eqref{eq:exch-torque} defines the effective stiffness that governs the entropic torque within the circular-magnon approximation.

Phenomenologically, the form of the torques in Eqs.~\eqref{eq:st1}, \eqref{eq:st2}, \eqref{eq:entropic1}, and \eqref{eq:entropic2} agrees with the form presented in Ref.~\cite{PhysRevLett.134.176401} after \(\boldsymbol\partial T\) is associated with the charge current \(\mathbf j_c\). For the \(d\)-wave altermagnet shown in Fig.~\ref{fig:altermagnet}, the symmetry of the response tensors implies \(\mathbf u\propto \hat{\sigma}_0\cdot \boldsymbol\partial T\) and \(\mathbf u^\prime\propto \hat{\sigma}_z\cdot \boldsymbol\partial T\), where \(\hat{\sigma}_0\) is the unit matrix and \(\hat{\sigma}_z\) is the Pauli matrix.

\textit{Linear response calculations.} -- 
For model calculations of the spin-splitter magnonic and entropic torques, we consider a square lattice model of altermagnet in Fig.~\ref{fig:altermagnet} with the Hamiltonian
\begin{align} \label{eq:Halt}
\mathcal H
&= J_1\!\!\sum_{\langle i,j\rangle}\!\mathbf{S}_{a i}\!\cdot\!\mathbf{S}_{b j}
 + \sum_{\langle i_x,j_x\rangle}\!\!\!\Big[J_2\,\mathbf{S}_{a i}\!\cdot\!\mathbf{S}_{a j}
+J_2'\,\mathbf{S}_{b i}\!\cdot\!\mathbf{S}_{b j}\Big]\\ \nonumber
&+ \sum_{\langle i_y,j_y\rangle}\!\!\!\Big[J_2'\,\mathbf{S}_{a i}\!\cdot\!\mathbf{S}_{a j}+J_2\,\mathbf{S}_{b i}\!\cdot\!\mathbf{S}_{b j}\Big]
 - K\sum_{i,\mu=a,b}\Big[(S_{\mu i}^z)^2\Big],
\end{align}
where  indices $a,b$ correspond to two sublattices.  
We perform the Holstein--Primakoff transformation to leading order in $1/S$.
The conventional coefficient $\mathbf u$ is obtained from the sum of the locally projected sublattice currents in Eq.~\eqref{eq:current}. Same-sublattice exchange gives a nonzero contribution, including in the reciprocal limit $J_2=J_2^\prime$. The entropic coefficient $\beta\mathbf u$ is calculated independently; $\beta$ remains a formal notation for comparison with Ref.~\cite{PhysRevLett.134.176401}.
For the spin-splitter response we choose the collinear reference $\mathbf S_A^{(0)}=-\hat{\mathbf z}$, $\mathbf S_B^{(0)}=+\hat{\mathbf z}$. With this spin-polarization convention, $\mathbf u^\prime=\mathbf j^s/s$ is determined by the spin Seebeck effect~\cite{PhysRevB.108.L180401}.
For AFM alignment, we can define the Nambu spinor 
describing magnons,
\(\Psi_{\vb k}=(\bm a_{\vb k},
\bm a^\dagger_{-\vb k})^T\), with \(\bm a_{\vb k}=(a_{\vb k,1},\dots,a_{\vb k,N})^{\mathsf T}.\)
We also introduce the Hamiltonian
\begin{equation}
H=\frac12\sum_{\vb k}\Psi_{\vb k}^\dagger \hat{H}_{\vb k}\Psi_{\vb k},\qquad
\hat{H}_{\vb k}=
\begin{pmatrix}
h_{\vb k} & \Delta_{\vb k}\\
\Delta_{\vb k}^\dagger & h_{-\vb k}^{\mathsf T}
\end{pmatrix}.
\end{equation}
The spin current is \( \mathbf{j}^s =\frac{\hbar}{4} \Psi^\dagger(\mathbf{r}) (\hat{\mathbf{v}}\hat{\sigma}_3 \hat{S}_z + \hat{S}_z\hat{\sigma}_3 \hat{\mathbf{v}}) \Psi(\mathbf{r}) \) where $\hat{\sigma}_3=\operatorname{diag}(I_N,-I_N)$ describes the bosonic metric and $\hat{\mathbf v}=\hbar^{-1}\partial_{\mathbf k}\hat H_{\mathbf k}$ is the velocity vertex. The dimensionless spin matrix is $\hat S_z=\operatorname{diag}(1,-1,1,-1)$ for the two-sublattice reference above~\cite{PhysRevLett.117.217203,PhysRevResearch.2.013079}. The paraunitary matrices $T_{\vb k}$ diagonalize the Hamiltonian, i.e., $T_{\vb k}^\dagger \hat{H}_{\vb k}T_{\vb k}=\sigma_3 E_{\vb k}$ with 
$E_{\vb k}=\big(\varepsilon_{1\vb k},\dots,\varepsilon_{N\vb k},
-\varepsilon_{1,-\vb k},\dots,-\varepsilon_{N,-\vb k}\big)$.
Defining the response coefficient as $j^{\,s}_i=-\alpha^s_{ij}\partial_j T$, we obtain using linear response approach~\cite{PhysRevB.101.024427} for the spin current tensor
\begin{equation}
\alpha^s_{ij}=\frac{k_B \hbar}{2 V}\sum_{n, \vb k}
\frac{\mathcal{S}_{nn}\,v_{ni}(\vb k)\,v_{nj}(\vb k)\tau_{n\vb k}}{E_{n\vb k}}
\Phi\!\left(\frac{E_{n\vb k}}{k_B T}\right),
\label{eq:spincurrent}
\end{equation}
where $V$ is the volume/area of the system, $\hat{\mathcal{S}}=\hat{\sigma}_3 T_{\vb k}^\dagger \hat{S}_z T_{\vb k}$~\cite{PhysRevB.101.024427}, $\vb v_n=\frac{1}{\hbar}\partial_{\vb k} E_{n\vb k}$, $\Phi(x)=x^{2}e^{x}/(e^{x}-1)^{2}$, and $\tau_{n\vb k}$ is the relaxation time. For an altermagnet with $d$-wave symmetry, we recover the shape of the response tensor $\hat{\alpha}^s =\alpha^s_0 \hat{\sigma_z}$, which agrees with symmetry analysis in Ref.~\cite{PhysRevLett.134.176401}. The form of the scattering time $\tau_{n\vb k}$ depends on the particular form of disorder. Starting with the stochastic LLG equation and taking all equilibrium
moments along $+\hat z$, we introduce canonical transverse amplitudes
$a_i=\sqrt{S/2}(\delta n_i^x+i\delta n_i^y)$, for which the particle equation is $i\hbar(1+i\alpha)\dot a_i=\partial H_2/\partial a_i^*+\eta_i$. For the corresponding Nambu
vector $\Psi_{\mathbf k}$, the quadratic energy becomes,
$H_2=\frac12\sum_{\mathbf k}\Psi_{\mathbf k}^\dagger
\hat{H}_{\vb k}\Psi_{\mathbf k}$, linearization gives
\begin{equation}
 i\hbar(\hat{\sigma}_3+i\alpha I)\partial_t\Psi_{\mathbf k}
 =\hat{H}_{\vb k}\Psi_{\mathbf k}+\eta_{\mathbf k},
 \label{eq:gilbert_nambu_supp}
\end{equation}
where $\eta_{\mathbf k}$ is the transformed thermal field.  Using the notation $\Psi_{\mathbf k}=T_{\mathbf k}\Gamma_{\mathbf k}$ and
multiplying by $T_{\mathbf k}^\dagger$, we obtain
\begin{equation}
 i\hbar[\hat{\sigma}_3+i\alpha T_{\mathbf k}^\dagger T_{\mathbf k}]\partial_t\Gamma_{\mathbf k}
 =\hat{\sigma}_3 E_{\vb k} \Gamma_{\mathbf k}+T_{\mathbf k}^\dagger\eta_{\mathbf k}.
 \label{eq:gilbert_band_supp}
\end{equation}
Within this approximation, the occupation relaxation time entering transport is
\begin{equation}
 \tau_{n\mathbf k}
 =\frac{\hbar}{2\alpha |E_{n\mathbf k}|
                    (T_{\mathbf k}^\dagger T_{\mathbf k})_{nn}}.
 \label{eq:occupation_lifetime_supp}
\end{equation}
The factor $(T_{\mathbf k}^\dagger T_{\mathbf k})_{nn}$ describes the antiferromagnetic enhancement of the linewidth~\cite{PhysRevB.110.134409}. 

Similarly, we obtain the gradient of the finite temperature effective exchange stiffness defined by the linear response equation 
$\partial_j \widetilde A^{A/B}_{ij}=\eta^{A/B}_{ij}\partial_j T$ with
\begin{equation}
\eta^{A/B}_{ij}=\frac{k_B  }{2 \mathcal{N}}\sum_{n, \vb k}
\frac{[\hat{\tilde{\mathcal{A}}}_{ij}^{A/B}]_{nn}}{E_{n\vb k}}
\Phi\!\left(\frac{E_{n\vb k}}{k_B T}\right),
\label{eq:sublattice}
\end{equation}
where $\mathcal{N}$ is the number of unit cells and $\hat{\tilde{\mathcal{A}}}^{A/B}_{ij}=\hat{\sigma}_3 T_{\vb k}^\dagger \hat{\tilde{A}}_{ij,\vb k}^{A/B} T_{\vb k}$. Here $\hat{\tilde{A}}_{ij,\vb k}^{A/B}$ corresponds to the finite temperature effective exchange stiffness in Eq.~\eqref{eq:exch-torque} expressed in matrix form in terms of \(\Psi_{\vb k}\) after applying the Holstein--Primakoff transformation. For the model in Fig.~\ref{fig:altermagnet}, we have $\sum_i[\hat{\tilde{A}}_{ii,\vb k}^{A}+\hat{\tilde{A}}_{ii,\vb k}^{B}]=-(J_1/2)\partial \hat{H}_{\vb k}/\partial J_1-J_2 \partial \hat{H}_{\vb k}/\partial J_2-J_2^\prime \partial \hat{H}_{\vb k}/\partial J_2^\prime-(2/S) \hat{I}_4(J_1-J_2-J_2^\prime)$ where $\hat{I}_4$ is a unit matrix.

The results of our calculations are shown in Fig.~\ref{fig:torques}, which includes the conventional magnonic coefficient $\alpha u/u_0$ together with the spin-splitter coefficient $\alpha u^\prime/u_0$ and the entropic coefficient $\beta u/u_0$. All three are positive for the stated reference and current convention. For the model in Eq.~\eqref{eq:Halt}, we use the parameters \(J_1=11.1~\mathrm{meV}\), \(J_2=-0.28~\mathrm{meV}\), \(J_2^\prime=-3.48~\mathrm{meV}\), and \(K=0.047~\mathrm{meV}\)~\cite{Gomonay2024}. The results in Fig.~\ref{fig:torques} indicate that the spin-splitter torque should provide a sizable contribution to the dynamics of magnetic textures. The curves are harmonic spin-wave results. Their upper-temperature portion is an extrapolation of this approximation, since thermal spin reduction is no longer small for $S=1/2$.
\begin{figure}
\centering
\includegraphics{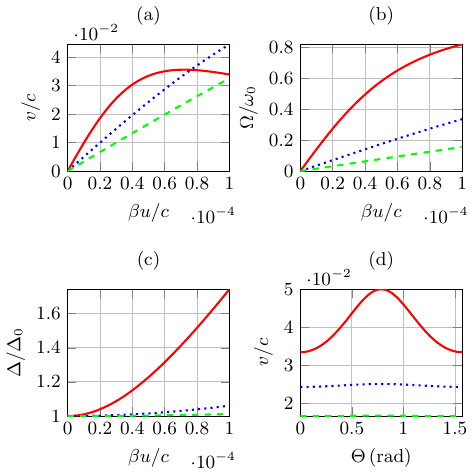}
\caption{(a) The domain wall velocity, (b) the angular precession speed $\Omega$, and (c) the domain wall width $\Delta$ as a function of the entropic drive $\beta u$ for $\Theta=0$. (d) The domain wall velocity for different directions of the temperature gradient with respect to the crystallographic axes, with $\beta u/c=5\times10^{-5}$.
The wall normal is parallel to $\boldsymbol\nabla T$ and rotates with it. All three torque coefficients are taken from Fig.~\ref{fig:torques} at $k_BT=30~\mathrm{meV}$: $\beta u=70.3\,\alpha u^\prime$ at $\Theta=0$, and $\alpha u=0.013\,\beta u$. The spin-splitter projection is $u^\prime(\Theta)=u^\prime(0)\cos2\Theta$, while $u$ and $\beta u$ are independent of $\Theta$.
The Gilbert damping is $\alpha=10^{-3}$ for the solid red, $\alpha=2\cdot10^{-3}$ for the dotted blue, and $\alpha=3\cdot10^{-3}$ for the dashed green curves.}
\label{fig:DW}
\end{figure}

\textit{Dynamics of magnetic textures.} -- 
Using the above calculated torques, we study the dynamics of domain walls in insulating altermagnets.  We use the Lagrangian written for the staggered field \cite{PhysRevLett.134.176401}:
\begin{equation}\label{eq:lag-l}
\begin{aligned}
L=\frac{ s}{2 \omega_\text{ex}}\int  \Big [(D_t\mathbf{n})^2 &-c^2(\partial_\alpha {\mathbf{n}})^2+\omega^2_0 n_z^2 \\
&+{\bf \boldsymbol{\mathcal{A}}}\cdot D_t\mathbf{n}+\boldsymbol{\mathcal{A}}_{wz}({\mathbf{u}}^\prime\cdot\boldsymbol{\partial}){\mathbf{n}} \Big ]d^2r,\\
\end{aligned}
\end{equation}
Here $D_t\mathbf n=\dot{\mathbf n}+(\mathbf u\cdot\boldsymbol\partial)\mathbf n$ includes the conventional adiabatic torque. We use its standard reactive completion, with the companion term $-(\mathbf u\cdot\boldsymbol\partial)\mathbf m$ in the magnetization equation, as in Ref.~\cite{PhysRevLett.134.176401}.
The vector $\bm{\mathcal{A}}=\Lambda\,\mathbf{n}\times\bigl(\partial_{x}^{2}\mathbf{n}-\partial_{y}^{2}\mathbf{n}\bigr)$ describes the altermagnetic interaction. In terms of the model in Fig.~\ref{fig:altermagnet}, we can write: $\omega_\text{ex}=16J_1/(s a_0^2)$,
$c=(4/a_0)\sqrt{J_1(J_1-J_2-J_2^\prime)/s^2}$, 
and $\omega_{0}=(8/a_0^2) \sqrt{J_1 K/s^2}$.
The magnonic spin-splitter torque is included by adding a term
containing the vector potential of the Wess--Zumino action
$\bm{\mathcal{A}}_{\mathrm{wz}}$, i.e.,
$\nabla_{\!\bm n}\times \bm{\mathcal{A}}_{\mathrm{wz}}=2 \omega_\text{ex}\mathbf{n}$.
The entropic torques in Eqs.~\eqref{eq:entropic1} and \eqref{eq:entropic2} are included via the Rayleigh density:
\begin{equation}
\mathcal{R}
= s
\left[
\frac{\alpha\,\dot{\mathbf{n}}^{2}}{2}
+ \,\dot{\mathbf{n}}\!\cdot\!(\beta\mathbf{u}\!\cdot\!\boldsymbol{\partial})\,\mathbf{n}+ \,\dot{\mathbf{m}}\!\cdot\!(\beta^\prime\mathbf{u}^\prime\!\cdot\!\boldsymbol{\partial})\,\mathbf{n}
\right],
\label{eq:Rfunction}
\end{equation}
where $\alpha$ is the Gilbert damping parameter. The ordinary time derivatives in $\mathcal R$ are measured in the laboratory frame and are not replaced by $D_t$. Note that the Euler-Lagrange-Rayleigh equation derived from Eq.~\eqref{eq:Rfunction} yields an additional higher-order torque term, $\boldsymbol \tau_m =  \mathbf{m} \times [\beta^\prime\mathbf{u}^\prime \cdot \boldsymbol{\partial}] \mathbf{n}$.
For convenience in what follows we use spherical coordinates, i.e., ${\bf n}=(\sin \theta \cos \phi,\sin \theta \sin \phi, \cos \theta)$. 

To study domain wall dynamics, we use the ansatz~\cite{Kosevich1990,Gomonay2024,PhysRevLett.134.176401}
$\cos\theta(x,t)=\tanh[(x-X(t))/\Delta(t)]$ and
$\phi(x,t)=\Phi(t)-b(t)[x-X(t)]/\Delta(t)$.
The collective variables $X(t)$ and $\Phi(t)$ describe the position and azimuthal angle, while $b$ and $\Delta$ adjust instantaneously in the approximation used here. Let $\hat{\mathbf e}$ be the wall normal, and write $u=\mathbf u\cdot\hat{\mathbf e}$, $u^\prime=\mathbf u^\prime\cdot\hat{\mathbf e}$, and $g=(\beta\mathbf u)\cdot\hat{\mathbf e}$. For a uniform, time-independent thermal drive, define
\begin{equation}
w=v-u,\qquad
Z_u=1-w^2/c^2-\Omega^2/\omega_0^2,
\end{equation}
where $v=\dot X$ and $\Omega=\dot\Phi$. Extremizing the conservative action with respect to the shape variables gives
\begin{equation}
\Delta=\Delta_0\,\frac{1-w^2/c^2}{\sqrt{Z_u}},
\qquad
b=\frac{\Delta w\Omega}{c^2-w^2},
\end{equation}
with $\Delta_0=c/\omega_0$ and $Z_u>0$. The equations of motion become
\begin{align}
\frac{\ddot X}{\omega_\text{ex}Z_u}
&=g-\alpha\dot X-bu^\prime,\label{eq:DW1}\\
\frac{\ddot\Phi}{\omega_\text{ex}Z_u}
&=\frac{u^\prime}{\Delta}(1+b^2)-\alpha\dot\Phi.\label{eq:DW2}
\end{align}
We neglect the same higher-order contribution proportional to $\beta^\prime u^\prime$ and the small altermagnetic correction to the wall shape. The damping terms depend on the laboratory velocity $v$, whereas $\Delta$, $b$, and $Z_u$ depend on $v-u$. In steady state these equations imply
\begin{equation}
v=\frac{g}{\alpha}\left(1-\frac{\Omega^2}{\omega_0^2}\right)
+u\,\frac{\Omega^2}{\omega_0^2}.
\end{equation}
Thus the conventional torque changes the speed of a precessing wall through its effect on the relative motion and the wall profile.

In Fig.~\ref{fig:DW}, we plot the results of Eqs.~\eqref{eq:DW1} and \eqref{eq:DW2} for different values of the Gilbert damping. In Fig.~\ref{fig:DW}(a) and (b), we observe that smaller Gilbert damping leads to faster domain-wall precession, which in turn slows the domain-wall motion. The precession originates from angular-momentum conservation: the angular momentum carried by magnons is absorbed by the wall, which causes precession, and then transferred to the lattice via Gilbert damping. As shown in Fig.~\ref{fig:DW}(c), the domain-wall width increases, consistent with the behavior reported in Ref.~\cite{Gomonay2024}, rather than undergoing the Lorentz contraction typical of antiferromagnets. The altermagnetic interaction $\Lambda$ can also affect domain-wall widths when $|v-u|\sim c$~\cite{Gomonay2024}, a regime not considered in our simulations.
In Fig.~\ref{fig:DW}(d), we observe anisotropy in the domain-wall speed for different temperature gradient directions relative to the crystallographic axes, parameterized by \(\Theta\). For $\Theta=\pi/4$, the spin-splitter current projected along the wall normal vanishes, so $\Omega=0$ and $v=g/\alpha$. The conventional torque still enters the wall width through $v-u$.

For the skyrmion examples below, we retain the approximation $u=0$ for the conventional adiabatic coefficient while keeping the independently specified entropic drive and spin-splitter torque. A finite $u$ would require recalculating the moving profile and dissipative tensor using the relative velocity $\mathbf v-\mathbf u$; however, we expect its effect to be small. To study skyrmion dynamics, we assume that the skyrmion texture is stabilized, for example, by an interfacial Dzyaloshinskii--Moriya interaction (DMI), which would require additional material engineering. For the model in Eq.~\eqref{eq:Halt}, we take the DMI strength to be $0.35\,\mathrm{mJ/m^2}$.
We assume a traveling-wave solution of the form ${\bf n}({\bf r}-{\bf v}t)$.
For velocities much smaller than $c$, we employ a $360^{\circ}$ domain-wall ansatz~\cite{Bttner2018,Wang2018,PhysRevB.50.16485}, while at larger velocities we use a deformed ``relativistic'' skyrmion profile~\cite{PhysRevB.106.L220402,SciPostPhys.8.6.086}.
The skyrmion position is parameterized by the generalized coordinate ${\bf X}(t)$, which yields the Thiele equation after substituting the skyrmion ansatz into the Lagrangian \eqref{eq:lag-l}~\cite{Gomonay2024}:
\begin{equation}\label{eq:sk-dynamics}
    \hat{\mathcal{M}}\,\ddot{\bf X}
    +\hat{\mathcal{G}}\,\dot{\bf X}
    +\alpha\,\hat{\mathcal{D}}\,\dot{\bf X}
    +\hat{\mathcal{B}}\,\dot{\bf X}
    ={\bf F},
\end{equation}
where $\hat{\mathcal{M}}$ is the skyrmion mass tensor, $\hat{\mathcal{G}}$ is the antisymmetric gyrotensor,
$\mathcal{D}_{ij}=\int d^2r\,(\partial_i{\bf n}\cdot\partial_j{\bf n})$ is the dissipative tensor, and
$\mathcal{B}_{ij}=-(1/2\omega_\text{ex})\int d^2r\,\big[\partial_i(\boldsymbol{\mathcal{A}}\cdot\partial_j{\bf n})-\partial_j(\boldsymbol{\mathcal{A}}\cdot\partial_i{\bf n})\big]$
arises from the altermagnetic term in the Lagrangian.
Here we consider a compensated antiferromagnet, for which the gyrotensor vanishes.
For the force we obtain
\begin{equation}
    {\bf F}=4\pi Q_{sk}\,\hat{\bf z}\times{\bf u}^\prime+\hat{\mathcal{D}}\cdot(\beta{\bf u}),
\end{equation}
where $Q_{sk}=(4\pi)^{-1}\int d^2r\,\mathbf n\cdot(\partial_x\mathbf n\times\partial_y\mathbf n)$ is the skyrmion charge, and we neglect the higher-order term proportional to $\beta^\prime{\bf u}^\prime$ that describes the anisotropic entropic torque.
Because of the spin-splitter magnonic torque, the force is not aligned with the temperature gradient, and we expect a temperature-gradient-induced skyrmion Hall effect.
The steady-state skyrmion velocity then reads
\begin{equation}\label{eq:velocity}
    {\bf v}=\frac{\beta {\bf u}}{\alpha}
    +\frac{4\pi Q_{sk}}{\alpha\,|\hat{\mathcal{D}}|}\,\hat{\bf z}\times\big(\hat{\mathcal{D}}\,{\bf u}^\prime\big),
\end{equation}
where $|\hat{\mathcal{D}}|=\det\hat{\mathcal{D}}$.
Above, for a smooth localized texture whose derivatives vanish at infinity, or under periodic boundary conditions we get $\hat{\mathcal{B}}=0$. The term $\hat{\mathcal{B}}$ might still be nonzero for open boundary conditions, e.g., for a nanostrip geometry~\cite{PhysRevLett.134.176401}.
\begin{figure}
\centering
\includegraphics{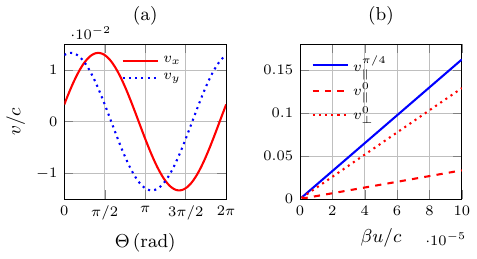}
\caption{Steady motion of an undeformed circular skyrmion with $Q_{sk}=+1$ and fixed dissipative tensor $\mathcal D_{ij}=15.392\,\delta_{ij}$ determined by relaxing a stationary skyrmion. (a) Cartesian velocity components $v_x$ and $v_y$ as the temperature-gradient direction rotates through the angle $\Theta$ relative to the crystallographic $x$ axis, at $\beta u/c=10^{-5}$. (b) Velocity components parallel and perpendicular to the temperature gradient as functions of $\beta u/c$, for $\Theta=0$ and $\Theta=\pi/4$. The vanishing curve $v_\perp^{\pi/4}=0$ is omitted. The torque coefficients are taken from Fig.~\ref{fig:torques} at $k_BT=30~\mathrm{meV}$, giving $\beta u=70.3\,\alpha u^\prime_0$, where $\mathbf u^\prime=u^\prime_0(\cos\Theta,-\sin\Theta)$. The Gilbert damping is $\alpha=3\times10^{-3}$. The static profile is held fixed at all drives; velocity-dependent deformation is neglected.}
\label{fig:skyrmion}
\end{figure}

In Fig.~\ref{fig:skyrmion}(a), we plot components of the skyrmion velocity, \(v_x\) and \(v_y\) from Eq.~\eqref{eq:velocity}, for different directions of the temperature gradient \(\boldsymbol{\partial}T\) with respect to the crystallographic axes, parameterized by \(\Theta\). The spin-splitter magnonic torque leads to the anisotropic skyrmion Hall effect. Interestingly, for the alignment corresponding to $\Theta=\pi/4$ we observe no side motion of skyrmion as can be seen in Fig.~\ref{fig:skyrmion}(b). This fast motion along the temperature gradient is mostly induced by the spin-splitter magnonic torque and can be useful for applications. 

\textit{Conclusions.} -- 
We have formulated a theory of thermomagnonic torques applicable to a general lattice model---for example, one describing an anisotropic magnet with exchange-driven magnon spin splitting---and applied it to insulating altermagnets. We have identified anisotropic entropic torques that drive magnetic textures toward the hot region under a temperature gradient, with the anisotropy arising from the reduced symmetry of the altermagnet. The conventional magnonic torque is also finite and modifies the coupled translational and precessional dynamics of a domain wall; however, for chosen parameters, its effect is relatively small. In addition, we identify a magnonic spin-splitter torque associated with sublattice-odd spin currents that are absent in conventional antiferromagnets, inducing domain-wall precession for certain crystal alignments and giving rise to an anisotropic skyrmion Hall effect under a temperature gradient. The latter can be particularly useful for racetrack memories, as it enables fast motion along a nanotrack driven by the magnonic spin-splitter torque without transverse deflection for appropriately engineered crystal alignments. Rare-earth orthoferrites such as $\mathrm{LuFeO_3}$ provide a possible setting for testing these effects. Quantitative velocity predictions require all three torque coefficients for the material-specific multisublattice model, evaluated with the occupation relaxation times; the linear-response framework is described in the Supplemental Material~\cite{Note}.
We note that for sharp magnetic textures, additional effects related to linear-momentum transfer need to be included~\cite{PhysRevB.90.104406,Shen2020JApplPhys}. Our results demonstrate that altermagnets exhibit certain properties of ferromagnets and ferrimagnets while retaining useful properties of conventional antiferromagnets. These findings can serve as a hallmark of altermagnetism in insulating altermagnets with magnetic textures.

\textit{Acknowledgments.} --
This work was supported by the U.S. Department of Energy, Office of Science, Basic Energy Sciences, under Award No. DE-SC0021019.

\bibliography{lib}
\end{document}


\title{Supplemental Material for ``Thermomagnonic Torques in Insulating Altermagnets"}

\author{Edward Schwartz}
\affiliation{Department of Physics and Astronomy and Nebraska Center for Materials and Nanoscience, University of Nebraska, Lincoln, Nebraska 68588, 
USA}
\author{Hamed Vakili}
\affiliation{Department of Physics and Astronomy and Nebraska Center for Materials and Nanoscience, University of Nebraska, Lincoln, Nebraska 68588, USA}
\author{Alexey A. Kovalev}
\affiliation{Department of Physics and Astronomy and Nebraska Center for Materials and Nanoscience, University of Nebraska, Lincoln, Nebraska 68588, USA}

\maketitle
\onecolumngrid

\begin{figure}
\centering
\includegraphics{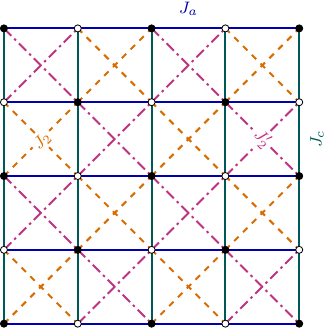}
  \caption{An effective spin model for the nonrelativistic part of the magnon spectrum of orthoferrite \(\mathrm{LuFeO_3}\). Filled circles denote the ``up'' sublattices and open circles denote the ``down'' sublattices of the collinear reference state. Blue solid lines mark $J_a$, teal solid lines mark $J_c$, orange dashed lines mark $J_{2}$, and magenta dash-dotted lines mark $J_{2}^\prime$. All bonds whose two endpoints lie inside the displayed window are drawn explicitly.}
  \label{fig:lufeo3}
\end{figure}

\section{Model}
We consider a minimal projected spin model motivated by the nonrelativistic magnon dispersion of orthorhombic \(\mathrm{LuFeO_3}\), omitting canting and relativistic mode mixing. The construction is motivated by two facts: (i) the experimentally relevant magnetic order contains four Fe sublattices, and (ii) the magnon splitting is strongest along the altermagnetic directions \(\Gamma\!-\!U\) and \(\Gamma\!-\!U'\), while it is absent along \(\Gamma\!-\!X\) and \(\Gamma\!-\!Z\). We therefore consider the following effective exchange model with easy-axis anisotropy in the projected \(a\)-\(c\) plane~\cite{GalindezRuales2025arXiv},
\begin{align}
H &= J_a \sum_{\langle ij\rangle_a} \mathbf{\hat{S}}_i\!\cdot\!\mathbf{\hat{S}}_j
   + J_c \sum_{\langle ij\rangle_c} \mathbf{\hat{S}}_i\!\cdot\!\mathbf{\hat{S}}_j
   - K_a \sum_i (\hat{S}_i^a)^2
\nonumber\\
&\quad + J_{2} \sum_{\langle ij\rangle_{14,+}} \mathbf{\hat{S}}_i\!\cdot\!\mathbf{\hat{S}}_j
      + J_{2}^\prime \sum_{\langle ij\rangle_{14,-}} \mathbf{\hat{S}}_i\!\cdot\!\mathbf{\hat{S}}_j
\nonumber\\
&\quad + J_{2}^\prime \sum_{\langle ij\rangle_{23,+}} \mathbf{\hat{S}}_i\!\cdot\!\mathbf{\hat{S}}_j
      + J_{2} \sum_{\langle ij\rangle_{23,-}} \mathbf{\hat{S}}_i\!\cdot\!\mathbf{\hat{S}}_j.
\label{eq:Hreal}
\end{align}
The hatted spins in Eq.~\eqref{eq:Hreal} are dimensionless quantum-spin operators of length $S$. This differs from the unit vectors used in the main-text Hamiltonian. The corresponding unit-vector exchange and anisotropy parameters are $S^2J$ and $S^2K_a$. Each site carries angular momentum $S\hbar$; for the four-site cell of area $v$ the total spin density is $s=4S\hbar/v$. The factors of $S$ in the spin-wave Hamiltonian below already implement this normalization.
For the plotted effective two-dimensional model we use equal projected cell lengths $a=c=a_0$ and $v=a_0^2$. Positions are given in cell units, and $k_a,k_c$ below denote dimensionless crystal momenta. 
Here the four sites are
\begin{equation}
\mathbf{r}_1=(0,0),\qquad
\mathbf{r}_2=\left(\frac12,0\right),\qquad
\mathbf{r}_3=\left(0,\frac12\right),\qquad
\mathbf{r}_4=\left(\frac12,\frac12\right),
\end{equation}
with collinear AFM  order
\(
1,4:\ \uparrow,\, 2,3:\ \downarrow
\), see Fig.~\ref{fig:lufeo3}.

The antiferromagnetic nearest-neighbor bonds are
\begin{align}
\langle ij\rangle_a &: (1,2),(4,3), \qquad \Delta \mathbf{r} = \pm\left(\frac12,0\right), \\
\langle ij\rangle_c &: (1,3),(4,2), \qquad \Delta \mathbf{r} = \pm\left(0,\frac12\right).
\end{align}
For the diagonal same-spin bonds the labels $+$ and $-$ mean displacements $\pm(1/2,1/2)$ and $\pm(1/2,-1/2)$, respectively.
We parametrize the corresponding couplings as
\begin{align}
J_{2}=J_{2,\mathrm{avg}}+ \delta_{\mathrm{alt}},\\
J_{2}^\prime=J_{2,\mathrm{avg}}- \delta_{\mathrm{alt}},
\end{align}
so that the exchange asymmetry changes sign between the two same-spin pairs. This gives a four-site conventional-cell representation of a $d$-wave altermagnetic exchange splitting. The reduced Hamiltonian also has the magnetic translation $(1/2,1/2)$, which exchanges $1\leftrightarrow4$ and $2\leftrightarrow3$; its four branches therefore include band folding.

\section{Linear spin-wave theory}

We rotate sublattices \(2\) and \(3\) by \(\pi\) so that all classical moments point along \(+\hat a\), and then perform a Holstein--Primakoff expansion,
\begin{equation}
S_i^a = S-a_i^\dagger a_i,
\qquad
S_i^+ \simeq \sqrt{2S}\, a_i,
\qquad
S_i^- \simeq \sqrt{2S}\, a_i^\dagger,
\end{equation}
keeping only the quadratic terms. In the Nambu basis
\begin{equation}
\Psi_{\mathbf{k}}=
\begin{pmatrix}
 a_{1,\mathbf{k}} & a_{2,\mathbf{k}} & a_{3,\mathbf{k}} & a_{4,\mathbf{k}} &
 a_{1,-\mathbf{k}}^{\dagger} & a_{2,-\mathbf{k}}^{\dagger} & a_{3,-\mathbf{k}}^{\dagger} & a_{4,-\mathbf{k}}^{\dagger}
\end{pmatrix}^{T},
\end{equation}
the quadratic Hamiltonian is
\begin{equation}
H_2=\frac12\sum_{\mathbf{k}}\Psi_{\mathbf{k}}^{\dagger}
\mathcal H_{\mathbf{k}}
\Psi_{\mathbf{k}},
\qquad
\mathcal H_{\mathbf{k}}=
\begin{pmatrix}
A_{\mathbf{k}} & B_{\mathbf{k}}\\
B_{\mathbf{k}}^{\dagger} & A_{\mathbf{k}}^{\ast}
\end{pmatrix}.
\label{eq:BdGform}
\end{equation}

Defining
\begin{equation}
c_a=\cos\frac{k_a}{2},\qquad c_c=\cos\frac{k_c}{2},
\qquad s_a=\sin\frac{k_a}{2},\qquad s_c=\sin\frac{k_c}{2},
\end{equation}
we obtain
\begin{equation}
A_{\mathbf{k}}=
\begin{pmatrix}
\varepsilon_0 & 0 & 0 & \alpha_- \\
0 & \varepsilon_0 & \alpha_+ & 0 \\
0 & \alpha_+ & \varepsilon_0 & 0 \\
\alpha_- & 0 & 0 & \varepsilon_0
\end{pmatrix},
\label{eq:Amatrix}
\end{equation}
with
\begin{align}
\varepsilon_0 &= 2S(K_a+J_a+J_c-2J_{2,\mathrm{avg}}), \\
\alpha_{\pm} &= 4S(J_{2,\mathrm{avg}} c_a c_c \pm \delta_{\mathrm{alt}} s_a s_c),
\end{align}
and
\begin{equation}
B_{\mathbf{k}}=2S
\begin{pmatrix}
0 & J_a c_a & J_c c_c & 0 \\
J_a c_a & 0 & 0 & J_c c_c \\
J_c c_c & 0 & 0 & J_a c_a \\
0 & J_c c_c & J_a c_a & 0
\end{pmatrix}.
\label{eq:Bmatrix}
\end{equation}

The magnon energies follow from the bosonic Bogoliubov problem
\begin{equation}
\Sigma_z\mathcal H_{\mathbf{k}}\,\mathbf{v}_{n\mathbf{k}}=E_{n\mathbf{k}}\,\mathbf{v}_{n\mathbf{k}},
\qquad
\Sigma_z=\operatorname{diag}(I_4,-I_4),
\label{eq:paraunitary}
\end{equation}
where $I_4$ is the identity matrix. The physical spectrum consists of four positive energies $E_{n\mathbf{k}}>0$, with angular frequencies $\omega_{n\mathbf{k}}=E_{n\mathbf{k}}/\hbar$. The Bogoliubov matrix is paraunitary: $T_{\mathbf{k}}^\dagger\Sigma_zT_{\mathbf{k}}=\Sigma_z$. Positive-energy eigenvectors are normalized by $\mathbf v_{n\mathbf k}^\dagger\Sigma_z\mathbf v_{n\mathbf k}=1$.

Equation~\eqref{eq:Amatrix} makes the symmetry transparent where the altermagnetic exchange enters only through the term
\begin{equation}
\pm 4S\,\delta_{\mathrm{alt}}\, s_a s_c
= \pm 4S\,\delta_{\mathrm{alt}}\sin\frac{k_a}{2}\sin\frac{k_c}{2},
\end{equation}
which vanishes identically on \(k_a=0\) and \(k_c=0\). Therefore the model is unsplit along \(\Gamma\!-\!Z\) and \(\Gamma\!-\!X\), while spin splitting appears away from those axes, i.e., along $\Gamma\!-\!U$ and $\Gamma\!-\!U'$. In this folded representation pairwise degeneracies also occur on the conventional-zone boundaries $k_a=\pm\pi$ or $k_c=\pm\pi$.

We use the following parameters
\begin{align}
S &= \frac52, &
J_a &= 17.9~\mathrm{meV}, &
J_c &= 10.3~\mathrm{meV},
\nonumber\\
J_{2,\mathrm{avg}} &= 1.3~\mathrm{meV}, &
\delta_{\mathrm{alt}} &= 0.9~\mathrm{meV}, &
K_a &= 0.015~\mathrm{meV}.
\label{eq:paramset}
\end{align}
Equivalently,
\(
J_{2}=2.2~\mathrm{meV},
\,
J_{2}^\prime=0.40~\mathrm{meV}
\).

\begin{figure}
\centering
\includegraphics{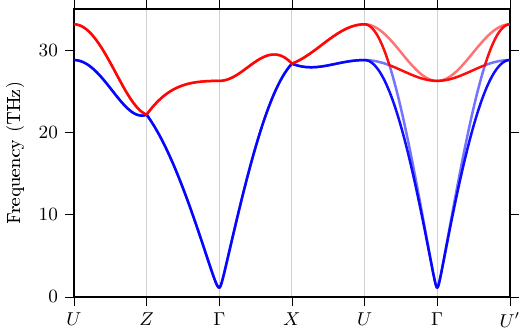}
\caption{Magnon spectrum of the projected model motivated by $\mathrm{LuFeO_3}$, using the phenomenological parameters in Eq.~\eqref{eq:paramset}. The plotted frequency is $E/h$. Colors distinguish energy-sorted branches and do not label magnon spin. Relativistic zone-center splitting and effects of canting are omitted. This spectrum approximately agrees with Fig.~1(c) of Ref.~\cite{GalindezRuales2025arXiv}.}
\label{fig:spectrum}
\end{figure}

The spectrum and all response curves below use the parameter set in Eq.~\eqref{eq:paramset}. These are phenomenological parameters of the projected model. In the positive-exchange convention of Eq.~\eqref{eq:Hreal}, both diagonal couplings are antiferromagnetic.
The high symmetry points are defined as
\begin{equation}
\Gamma=(0,0),\qquad X=(\pi,0),\qquad Z=(0,\pi),\qquad U=(\pi,\pi),\qquad U'=(-\pi,\pi).
\end{equation}
In Fig.~\ref{fig:spectrum}, we plot all four positive branches as frequencies $f_{n\mathbf k}=E_{n\mathbf k}/h$ in THz, with $h=2\pi\hbar$.

\begin{figure}
\centering
\includegraphics{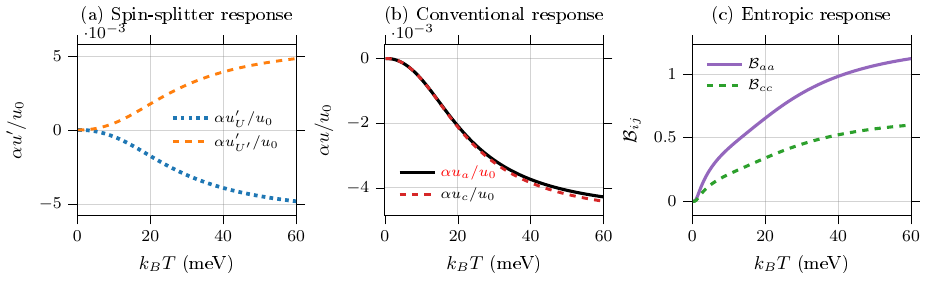}
    \caption{Thermomagnonic response of the four-site model used in Fig.~\ref{fig:spectrum}. (a) The longitudinal spin-splitter responses for gradients along $U$ and $U'$. (b) Conventional spin-transfer coefficients for gradients along $a$ and $c$, using the same $\mathbf u$ definition as in the main text. (c) Entropic responses for the crystallographic axes, $a=X$ and $c=Z$. All curves use $S=5/2$, $s=4S\hbar/v$, and $u_0=(k_B/s)|\boldsymbol\nabla T|$.}
    \label{fig:JaJc_python_style_twopanel}
\end{figure}

\section{Linear response calculations}

For the plotted effective two-dimensional model, we use equal projected cell
lengths $a=c=a_0$ and cell area $v=a_0^2$.  The momenta $k_a,k_c$ used above
are dimensionless, with physical wave vectors $k_a/a_0,k_c/a_0$ and
$k_a,k_c\in[-\pi,\pi)$.  Thus $X$ and $Z$ denote the $a$ and $c$ axes,
while $\hat{\mathbf e}_U=(\hat{\mathbf e}_a+\hat{\mathbf e}_c)/\sqrt2$ and
$\hat{\mathbf e}_{U'}=(-\hat{\mathbf e}_a+\hat{\mathbf e}_c)/\sqrt2$.

We group the down sites into $A=\{2,3\}$ and the up sites into
$B=\{1,4\}$. In terms of quantum-spin
fluctuations and the unit reference directions $\mathbf n_\ell^{(0)}$,
the site currents corresponding to Eq.~(11) of the main text are
\[
 \mathbf j_\ell=-\frac1v\sum_{\boldsymbol\rho}
 \boldsymbol\rho J_{\ell,\boldsymbol\rho}\,
 \mathbf n_\ell^{(0)}\cdot
 \left\langle\hat{\mathbf S}_\ell^\perp\times
 \hat{\mathbf S}_{\ell+\boldsymbol\rho}^\perp\right\rangle.
\]
The grouped currents are sums,
$\mathbf j_A=\mathbf j_2+\mathbf j_3$ and
$\mathbf j_B=\mathbf j_1+\mathbf j_4$, and
\[
\mathbf u=-\frac{\mathbf j_A+\mathbf j_B}{2s},\qquad
\mathbf u'=\frac{\mathbf j_A-\mathbf j_B}{2s}
           =\frac{\mathbf j^{\mathrm s}_a}{s}.
\]
Here $\mathbf j^{\mathrm s}_a$ is the physical angular-momentum current
polarized along $+\hat a$.  

For $u_0=(k_B/s)|\boldsymbol\nabla T|$ and
$\hat{\mathbf g}=\boldsymbol\nabla T/|\boldsymbol\nabla T|$, we introduce tensors $\mathcal U_{ij}$, $\mathcal P_{ij}$, and $\mathcal B_{ij}$, parametrizing linear response
\[
 \frac{\alpha u_i}{u_0}=\mathcal U_{ij}\hat g_j,\qquad
 \frac{\alpha u_i'}{u_0}=\mathcal P_{ij}\hat g_j,\qquad
 \frac{[\beta u]_i}{u_0}=\mathcal B_{ij}\hat g_j.
\]

For the entropic response, we introduce sum of the four site contributions to the effective stiffness in
Eq.~(13) of the main text, $\widetilde A^{\mathrm{tot}}_{ij}$, i.e.,
$[\beta u]_i=-(2s)^{-1}\partial_j\widetilde A^{\mathrm{tot}}_{ij}$.

After Brillouin-zone integration, symmetry gives
$\mathcal U_{ac}=\mathcal U_{ca}=\mathcal B_{ac}=\mathcal B_{ca}=0$ and
$\mathcal P_{aa}=\mathcal P_{cc}=0$, with
$\mathcal P_{ac}=\mathcal P_{ca}$.
Consequently the longitudinal spin responses are
$\mathcal P_{UU}=\mathcal P_{ac}$ and
$\mathcal P_{U'U'}=-\mathcal P_{ac}$.
For the declared positive same-spin exchanges, the calculated
$\mathcal U_{aa}$ and $\mathcal U_{cc}$ are negative, as is
$\mathcal P_{UU}$, whereas $\mathcal P_{U'U'}$ is positive.
Figure~\ref{fig:JaJc_python_style_twopanel} shows these responses. The entropic response is positive and anisotropic, and all thermal responses vanish as $T\to0$.
The calculation uses a fixed collinear background and harmonic magnons. At $k_BT=60~\mathrm{meV}$ the thermal spin reduction per site is approximately $0.42S$, so the upper-temperature part is a harmonic extrapolation. 

\bibliography{lib}